# Collapse of ultra-short pulse of electromagnetic field within non-linear electrodynamics

M.B. Belonenko, N.N. Konobeeva

Volgograd State University

mbelonenko@yandex.ru

We analyze the development of pulse instability within non-linear electrodynamics based on the Maxwell's equations, without the approximation of slowly varying amplitudes and phases. The action is determined on the basis of the logarithmic and exponential Lagrangians, built on the invariants of the electromagnetic field. It is shown that the onset of the collapse of an ultra-short pulse of the electromagnetic field in the framework of nonlinear electrodynamics.

## 1. Intriduction

Problems related to the consideration of nonlinear electrodynamics (NE) in vacuum have recently attracted a great attention from researchers. Now, NE effects associated with consideration of the Heisenberg-Euler Lagrangian [1,2], which is generally given as an integral, are well studied, and consideration is carried out using expansion in a series of scalar and pseudoscalar invariants of the electromagnetic field. The classical Lagrangian of the electromagnetic field is a member of the first order and contains only one invariant (scalar). Among the studies in the framework of this approach, we note Ref. [3–6], as well as Ref. [7, 8], which consider the effects of subsequent orders. From the point of view of quantum electrodynamics, such nonlinearities arise from the exchange of virtual electron-positron pairs between photons in a vacuum. Note also that this approach is easily generalized to the case of "rapidly changing" fields, as described in [9, 10]. In this approach, all the terms and the Heisenberg-Euler Lagrangian itself is directly obtained from the classical Lagrangian of quantum electrodynamics.

At the same time, a different approach has historically been formed. It began to develop in consequence of the fact that in classical electrodynamics the field of a point charge tends to infinity as it approaches a point charge. To get rid of this circumstance, a number of authors proposed nonlinear Lagrangians, which give the usual electrodynamic effects in the case of weak fields, but in consequence of nonlinearity, give a finite value of the field of a point charge at the origin of coordinates, coinciding with the charge. There is a lot of such Lagrangians. And, the choice is largely determined by the preferences of the authors. In this paper, we chose the Lagrangians proposed in [11–13]. To test this hypothesis, it is proposed to use the propagation of ultra-short pulses in a vacuum. In the case of conventional electrodynamics, such pulses should spread (due to dispersion). The question of the stability of such pulses in the case of NE is still open.

We use the direct solution of the Maxwell's equations without the he slowly varying envelope approximation (SVEA). As is well known, the SVEA has several disadvantages, such as changing the dispersion law of linear oscillations in the Maxwell system of equations before and after applying the SVEA. Also after applying the SVEA, both spatial and temporal dispersions of nonlinear terms are not taken into account.

**2. Basic equations.**

The Maxwell's system of equations (with explicit **D** and **H vectors in terms of E and B vectors**) has the form:

$$\nabla \cdot \mathbf{E} = (\rho - \nabla \cdot \mathbf{P})/\varepsilon_0,$$
$$\nabla \cdot \mathbf{B} = 0,$$
$$\frac{\partial \mathbf{B}}{\partial t} + \nabla \times \mathbf{E} = 0, \qquad (1)$$
$$\frac{1}{c^2}\frac{\partial \mathbf{E}}{\partial t} - \nabla \times \mathbf{E} = -\mu_0 \left( \mathbf{j} + \frac{\partial \mathbf{P}}{\partial t} + \nabla \times \mathbf{M} \right)$$

This system, in the absence of free charges and currents, can be written as:

$$\frac{1}{c^2}\frac{\partial^2 \mathbf{E}}{\partial t^2} - \nabla^2 \mathbf{E} = -\mu_0 \left( \frac{\partial^2 \mathbf{P}}{\partial t^2} + c^2 \nabla (\nabla \cdot \mathbf{P}) + \frac{\partial}{\partial t}(\nabla \times \mathbf{M}) \right),$$
$$\frac{1}{c^2}\frac{\partial^2 \mathbf{B}}{\partial t^2} - \nabla^2 \mathbf{B} = \mu_0 \left( \nabla \times (\nabla \times \mathbf{M}) + \frac{\partial}{\partial t}(\nabla \times \mathbf{P}) \right) \qquad (2)$$

To establish the connection of vectors **M** and **P** with **E** and **B** we use logarithmic and exponential Lagrangians in the form, as suggested by Soleng (3) [11] and Handy (4) [12, 13] respectively:

$$L_{LN} = -4\beta^2 \ln\left(1 + \frac{F^2}{4\beta^2}\right) \qquad (3)$$

$$L_{EXP} = 4\beta^2 \left( exp\left(-\frac{F^2}{4\beta^2}\right) - 1 \right) \qquad (4)$$

where $\beta$ is the nonlinear parameter with the dimension of mass, and the pseudoscalar invariant of the electromagnetic field is determined by the ratio:

$$F = \sqrt{E^2 - c^2 B^2} \qquad (5)$$

The interaction term in Lagrangian has the form:

$$L = L_{LN/EXP} + F^2 \qquad (6)$$

Taking into account that $P = \delta L/\delta E$, $M = \delta L/\delta B$ it is easy to obtain expressions for magnetization and polarization for logarithmic (7a) and exponential (7b) Lagrangians:

$$P = 2E\left(1 - \frac{4\beta^2}{4\beta^2 + F^2}\right), \quad M = -2B\left(1 - \frac{4\beta^2}{4\beta^2 + F^2}\right) \tag{7a}$$

$$P = 2E\left(1 - exp\left(-\frac{F^2}{4\beta^2}\right)\right), \quad M = -2c^2 B\left(1 - exp\left(-\frac{F^2}{4\beta^2}\right)\right) \tag{7b}$$

### 3. Main results of numerical simulation

We chose initial conditions for the equation system (2, 7) in the form:

$$\begin{aligned}
&E_z = 0, \\
&B_x = B_y = B_z = 0, \\
&E_x = E_y = A \cdot exp\left(-\left(\frac{z}{\gamma}\right)^2\right) exp\left(-\frac{x^2 + y^2}{\gamma_p^2}\right), \\
&\frac{d}{dt}E_z = \frac{d}{dt}B_x = \frac{d}{dt}B_y = \frac{d}{dt}B_z = 0, \\
&\frac{d}{dt}E_x = \frac{d}{dt}E_y = \frac{2vA}{\gamma^2} \cdot exp\left(-\left(\frac{z}{\gamma}\right)^2\right) exp\left(-\frac{x^2 + y^2}{\gamma_p^2}\right)
\end{aligned} \tag{8}$$

It corresponds to the propagation of a cylindrically symmetric electric field pulse of the Gaussian form at the initial moment. In eq. (8) $v$ is the initial velocity of the pulse, $\gamma$ is the pulse width along the propagation direction, $\gamma_p$ is the pulse width in the transverse direction. Further, we use the cylindrical coordinate system:

$$\begin{aligned}
&\Delta \bar{B} = \left(\Delta B_\rho - \frac{B_\rho}{\rho^2}\right)\hat{\rho} + \left(\Delta B_\varphi - \frac{B_\varphi}{\rho^2}\right)\hat{\varphi} + \Delta B_z \cdot \hat{z}, \\
&\bar{\nabla} \times \bar{B} = -\frac{\partial B_\varphi}{\partial z}\hat{\rho} + \left(\frac{\partial B_\rho}{\partial z} - \frac{\partial B_z}{\partial \rho}\right)\hat{\varphi} + \frac{1}{\rho}\frac{\partial(\rho B_\varphi)}{\partial \rho}\hat{z}, \\
&\bar{\nabla} \cdot \bar{B} = \frac{1}{\rho}\frac{\partial(\rho B_\rho)}{\partial \rho} + \frac{\partial B_z}{\partial z}, \\
&\Delta f = \frac{1}{\rho}\left(\frac{\partial}{\partial \rho}\left(\rho \frac{\partial f}{\partial \rho}\right)\right) + \frac{\partial^2 f}{\partial z^2}, \\
&\bar{\nabla} f = \frac{\partial f}{\partial \rho} \cdot \hat{\rho} + \frac{\partial f}{\partial z} \cdot \hat{z}, \\
&\rho = \sqrt{x^2 + y^2}, \quad tg\varphi = \frac{y}{x}
\end{aligned} \tag{9}$$

and we apply the cross type numerical scheme [14].

The typical pulse evolution for Lagrangian (3) is shown in Fig. 1, and for Lagrangian (4) in Fig. 2. And, as usual, $E_\rho = E_x \cdot cos\varphi + E_y \cdot sin\varphi$.

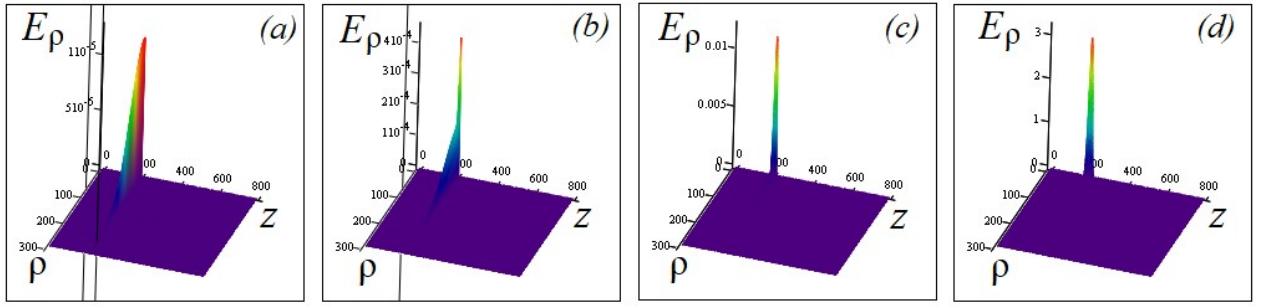

Fig. 1. The dependence of the components of the electric field strength $E_\rho$ on the coordinates at different points in time for the Lagrangian (3): (*a*) *t*=0; (*b*) *t*=100; (*c*) *t*=300; (*d*) *t*=620. All values are in the relative units.

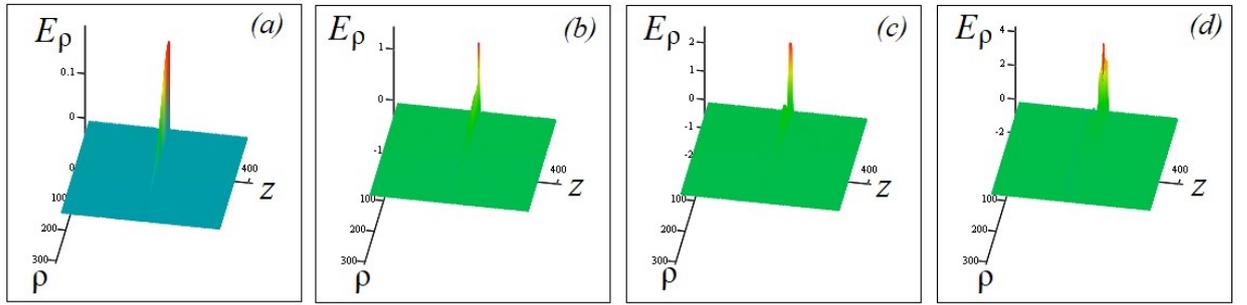

Fig. 2. The dependence of the components of the electric field strength $E_\rho$ on the coordinates at different points in time for the Lagrangian (4): (*a*) *t*=0; (*b*) *t*=100; (*c*) *t*=200; (*d*) *t*=430. All values are in the relative units.

When specifying the initial conditions, it was assumed that the initial velocity is u=0999·c, where c is the light speed. This corresponds to the fact that the pulse enters the vacuum from a medium with a refractive index not much greater than 1.

At the first stage of evolution, at the front of pulse there is a split into a series of smaller pulses, which, naturally, is accompanied by diffraction spreading in the perpendicular direction. This effect is not detected earlier, since the variational method most frequently used for analysis contains an assumption about the preservation of the pulse shape (that is, only the parameters describing this form change over time). At the second stage, we see the growth of the pulse amplitude at the front, i.e. his collapse. It should be noted, that the time after which the second stage begins depends most strongly on the amplitude of the pulse at the initial moment of time. It is important here that when we investigate such powerful ultra-short pulses, pretty quickly the pulse will no longer satisfy the condition: $|E| \ll E_{crit}, E_{crit} = \frac{m_e c^2}{e \lambda_e} \sim 10^{18}$ V/m and the birth of electron-positron pairs begins, which is not taken into account in our model.

Amplitude slices $E_\rho$ on the z axis for the Lagrangian (4) are presented in Fig.3.

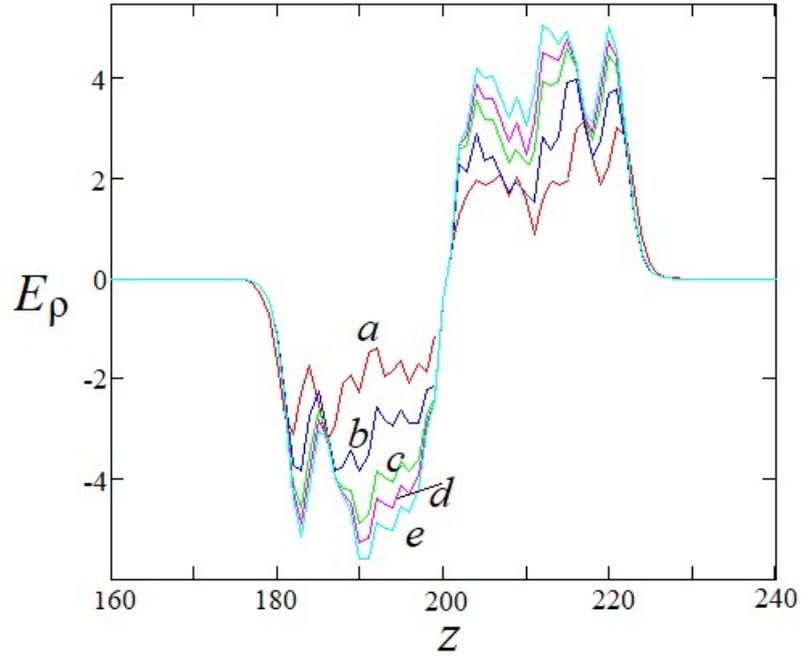

Fig. 3. The dependence of the components of the electric field strength $E_\rho$ for longitudinal section from β at t=420: a) $\beta=7$; b) $\beta=13$; c) $\beta=19$; d) $\beta=22$; e) $\beta=25$. All values are in the relative units.

The results shown in Fig. 3 confirm the fragmentation of the pulse, and allow us to conclude that the collapse can be accelerated by increasing the nonlinear parameter β.

Thus, it is possible to analyze the mechanism of instability arising from the propagation of a powerful Gaussian pulse in a vacuum. Note that in Ref. [5], only the estimate obtained by the variational method and using the SVEA for the effective pulse width is given, and a conclusion is also drawn about the collapse.

## 4. Conclusion

In conclusion, we formulate our main results:

1) The solution of the Maxwell's equations is obtained for two types of nonlinear electrodynamics, which eliminate the divergence of the electric field at r=0.

2) The scenario of the onset of the collapse of the ultra-short pulse of the electromagnetic field within the framework of nonlinear electrodynamics is described. This scenario includes a rapid increase in the pulse in the direction of propagation in a finite period of time.

## 5. Acknowledgment

This work was supported by the Ministry of Science and Higher Education of the Russian Federation (state assignment no. 2019-0730 "Supercomputer modeling of the dynamics of continuous media").